\documentclass[conference]{IEEEtran}
\IEEEoverridecommandlockouts
\usepackage{cite}
\usepackage{amsmath,amssymb,amsfonts,latexsym}
\usepackage{braket}
\usepackage{algorithmic}
\usepackage{graphicx}
\usepackage{textcomp}
\usepackage{xcolor}
\usepackage{euscript,mathrsfs}
\usepackage{mathtools}
\usepackage{yhmath}
\usepackage{url}
\usepackage{empheq}
\usepackage{booktabs}
\usepackage{hyperref}
\usepackage{lipsum}
\usepackage[all]{xy}
\usepackage{tikz-cd}

\def \vt{\vartheta}

\def \s{\sigma}
\def \p{\partial}

\DeclareMathOperator*{\argmin}{arg\,min}

\def\BibTeX{{\rm B\kern-.05em{\sc i\kern-.025em b}\kern-.08em
    T\kern-.1667em\lower.7ex\hbox{E}\kern-.125emX}}

\makeatletter
\newcommand{\linebreakand}{%
  \end{@IEEEauthorhalign}
  \hfill\mbox{}\par
  \mbox{}\hfill\begin{@IEEEauthorhalign}
}
\makeatother

\begin{document}

\title{Quantum Gradient Class Activation Map for Model Interpretability\\
\thanks{The views expressed in this article are those of the authors and do not represent the views of Wells Fargo. This article is for informational purposes only. Nothing contained in this article should be construed as investment advice. Wells Fargo makes no express or implied warranties and expressly disclaims all legal, tax, and accounting implications related to this article.}
}



\author{
\IEEEauthorblockN{
    Hsin-Yi Lin \IEEEauthorrefmark{1}\IEEEauthorrefmark{2},
    Huan-Hsin Tseng \IEEEauthorrefmark{3}\IEEEauthorrefmark{4},
    Samuel Yen-Chi Chen\IEEEauthorrefmark{5}\IEEEauthorrefmark{6}, Shinjae Yoo \IEEEauthorrefmark{3}\IEEEauthorrefmark{7}
}
\IEEEauthorblockA{\IEEEauthorrefmark{1} Department of Mathematics and Computer Science, Seton Hall University, South Orange NJ, USA}
\IEEEauthorblockA{\IEEEauthorrefmark{3}  Artificial Intelligence 
 \& Machine Learning Department, Brookhaven National Laboratory, Upton NY, USA}
\IEEEauthorblockA{\IEEEauthorrefmark{5} Wells Fargo, New York, NY, USA}

\IEEEauthorblockA{Email:
\IEEEauthorrefmark{2}
hsinyi.lin@shu.edu,
\IEEEauthorrefmark{4}
htseng@bnl.gov,
\IEEEauthorrefmark{6}
yen-chi.chen@wellsfargo.com,
\IEEEauthorrefmark{7}
syjoo@bnl.gov
}}

\maketitle

\begin{abstract}
Quantum machine learning (QML) has recently made significant advancements in various topics. Despite the successes, the safety and interpretability of QML applications have not been thoroughly investigated. This work proposes using Variational Quantum Circuits (VQCs) for activation mapping to enhance model transparency, introducing the Quantum Gradient Class Activation Map (QGrad-CAM). This hybrid quantum-classical computing framework leverages both quantum and classical strengths and gives access to the derivation of an explicit formula of feature map importance. Experimental results demonstrate significant, fine-grained, class-discriminative visual explanations generated across both image and speech datasets.



\end{abstract}


\section{\label{sec:Indroduction}Introduction}
%
Recent advances in quantum computing and machine learning (ML) have drawn significant attention, leading to efforts to combine these two fascinating technologies. Although current quantum devices still face challenges due to noise and other imperfections, a hybrid quantum-classical computing framework has been proposed to leverage the strengths of both quantum and classical computing \cite{bharti2022noisy}. Variational quantum circuits (VQCs) serve as the foundational elements of this hybrid framework. Within this framework, computational tasks that can benefit from quantum advantages are executed on quantum computers, while others are handled by classical computers. VQC-based algorithms have been shown to have certain advantages over classical models \cite{abbas2021power,du2020expressive,caro2022generalization} and have demonstrated success in various ML tasks, including classification \cite{mitarai2018quantum,chen2022quantumCNN,cong2019quantum}, sequential modeling \cite{chen2022quantumLSTM}, audio and language processing \cite{di2022dawn,stein2023applying,li2023pqlm}, and reinforcement learning \cite{chen2020variational,lockwood2020reinforcement,yun2023quantum}.

Despite these successes, certain aspects of quantum machine learning (QML) models have not been thoroughly investigated, particularly concerning the safety of QML applications.
Model interpretability and transparency are crucial for understanding and ensuring proper use, especially due to extensive machine learning applications and the continuing development of policy and regulation. With the rapid development of quantum computing, it is important to explore whether any quantum advantage could be offered in this direction.



Various approaches have been developed to address model interpretability from different aspects, such as from feature importance techniques \cite{fisher2019all} and rule-based methods \cite{nori2021accuracy}. Model-agnostic methods  
provide local interpretability by approximating complex models with simpler ones \cite{ribeiro2016should,lundberg2017unified,ribeiro2018anchors}.

Among diverse approaches, visual techniques stand out for their intuitive appeal. Methods such as Partial Dependence Plot (PDP)\cite{friedman20011999} and
Individual Conditional Expectation (ICE) \cite{goldstein2015peeking} visualize the relationship between a feature and the predicted outcome. Class Activation Mapping (CAM) \cite{zhou2016learning} and its variants \cite{selvaraju2017grad, chattopadhay2018grad} have emerged as powerful tools for generating class-discriminative localization as visual explanations for Convolutional Neural Networks (CNNs).

In this work, we propose using VQC for activation mapping as a pioneering exploration of quantum circuit applications for model transparency.  Our proposed method, Quantum Gradient Class Activation Map (QGrad-CAM), employs a VQC to weigh the importance of activation maps generated from a CNN-based network. 


Based on the structure of VQC, we will derive the explicit formula for the importance of each activation map, which can serve as an example of this advantage. Furthermore, experiments are performed on image and speech datasets for validation. Meaningful highlighted regions are returned using the proposed method for all cases.

To summarize, our main contributions include
\begin{itemize}
     \item Deriving an explicit formula for the VQC importance of activation maps and visualize model decisions.
     \item Conducting experiments to show the VQC can effectively weigh the importance of activation maps and provide textual explanations for model decisions.
 \end{itemize}

The rest of the paper is organized as follows. In Sec.~\ref{sec:Related_Work}, we point out relevant work in previous literature. Our proposed method, QGrad-CAM, is introduced in Sec.~\ref{sec:QGRAD_CAM}. The details of QGrad-CAM are written in Sec.~\ref{sec:QGRAD_CAM_with_VQC} after a VQC review and discussions of the critical ideas in Sec.~\ref{sec:VQC} and Sec.~\ref{sec:Regularities_VQC} that motivate our proposal. With the VQC structure, we derive the explicit formula for the importance of feature maps in Sec.~\ref{sec:Explainability_QGRAD_CAM}. The experimental results can be found in Sec.~\ref{sec:Experiment}, and the conclusions in Sec.~\ref{sec:Conclusion}.

\section{\label{sec:Related_Work}Related work}
\textbf{Explaining QML Models.}
VQC-based QML methods have demonstrated significant success in the domain of classification \cite{mitarai2018quantum}. Noteworthy advancements include the development of Quantum Convolutional Neural Networks (QCNN) \cite{cong2019quantum, chen2022quantumCNN}, quantum transfer learning techniques \cite{mari2020transfer}, and hybrid models incorporating tensor networks \cite{chen2021end}. Despite these achievements, the explainability of these models has not been thoroughly addressed.
%
Several preliminary attempts have been made to explain the predictions generated by QML models. For instance, the study in \cite{power2024feature} investigates feature importance within quantum Support Vector Machine (SVM) models using the Iris dataset, which consists of only four features. The generalizability of this method to larger-scale datasets, such as image data or hybrid models combining classical NNs and VQCs, remains uncertain. Another approach, as outlined in \cite{heese2023explaining}, involves calculating the Shapley values for gates in VQCs to determine their respective contributions. The objective of the methods presented in \cite{heese2023explaining} is to rigorously evaluate the quality of various circuit architectures. In contrast, our proposed framework aims to uncover feature importance given a fixed VQC architecture using a scalable gradient-based method.
%

\textbf{Visualizing \& CNN localization.}
CNNs have long demonstrated exceptional performance across a wide range of applications. This sparked extensive research aimed at understanding the underlying properties of CNNs. Several works developed techniques for visualizing the CNN-learned latent representation by, for example, analyzing convolution layers \cite{zeiler2014visualizing, zhou2014object} and inverting deep features \cite{dosovitskiy2015inverting, mahendran2015understanding, mahendran2016visualizing}. The research prompted the discovery of CNNs' ability to localize objects. CAM was proposed in \cite{zhou2016learning}, where CNN layers localize objects unsupervised and produce visual explanations for each class. Different pooling methods were explored in \cite{pinheiro2015image, oquab2015object} with a similar structure as CAM.
Grad-CAM \cite{selvaraju2017grad} generalized CAM by combining the class discriminative property with gradient techniques. The method allowed fine-grained discriminative localization and was improved, especially for multiple instances scenarios in \cite{chattopadhay2018grad}.




\section{\label{sec:QGRAD_CAM}Quantum gradient class-activation map (QGrad-CAM)}

\subsection{\label{sec:VQC} Variational Quantum Circuits}
\begin{figure}[htb]
\begin{center}
\begin{tikzcd}
\mathcal{H} \arrow[r, "U(\theta)"] & \mathcal{H} \arrow[d, "<\cdot | Q | \cdot>"]\\
\mathbb{R}^n \arrow[u, "V(x_j)"] \arrow[r, dashed, "f_{\text{VQC}}"]
& \mathbb{R}^m
\end{tikzcd}
\end{center}
\caption{The diagram of a VQC with a classical input $x_j \in \mathbb{R}^n$ received.}\label{fig: VQC}
\end{figure}
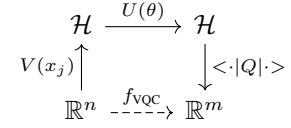

VQCs also referred to as Parameterized Quantum Circuits (PQCs), are quantum circuits characterized by tunable parameters that can be optimized based on specific metrics or signals \cite{mitarai2018quantum}. VQCs or PQCs serve as fundamental components of Quantum Neural Networks (QNNs), which underlie current QML techniques. The VQC as a QML method operates $n$ qubits in space $\mathcal{H} = \bigotimes^n \mathbb{C}^2 \cong \mathbb{C}^{2^n}$, where $\mathcal{H}$ is a Hilbert space containing a standard basis written as $\beta = \{\ket{00\cdots 0}, \ket{00\cdots 1}, \ldots, \ket{11\cdots 1} \}$ such that any quantum state $\ket{\psi} \in \mathcal{H}$ can be expanded by $\beta$, \emph{i.e.,}
\[
\ket{\psi} = c_0 \ket{00\cdots 0} + \cdots + c_N \ket{11\cdots 1}
\]
for some coefficients $c_i \in \mathbb{C}$ with $N = 2^n$. Let $\mathcal{L}(\mathcal{H})$ denotes all linear operators on $\mathcal{H}$ and $\mathcal{U}(\mathcal{H})$ be the collection of unitary operators in $\mathcal{L}(\mathcal{H})$. A VQC typically functions through the following three steps (see Fig.~\ref{fig: VQC}),
\begin{enumerate}
    \item A quantum encoding $V:\mathbb{R}^n \to \mathcal{U}(\mathcal{H})$,
    \item A \emph{variational} quantum gate $ U(\theta) \in \mathcal{U}(\mathcal{H})$ parameterized by $\theta$,
    \item A quantum measurement of $Q$ as an output $\langle \cdot | Q | \cdot \rangle: \mathcal{H} \to \mathbb{R}$.
\end{enumerate}

Assume data is of classical form $\mathcal{D} = \{(x_j, y_j) \, | \, x_j \in \mathbb{R}^n, y_j \in \mathbb{R}^m, \, j=1,\ldots, N \}$ where $x_j$ is an input of sample index $j$ and $y_j$ be the corresponding label. A quantum encoding scheme chooses a fixed gate sequence $V:\mathbb{R}^n \to \mathcal{U}(\mathcal{H})$ to convert classical data into quantum states such that each input corresponds to a unitary, $x_j \mapsto V(x_j)$. One then associates the quantum state $\ket{\psi_{x_j}}:= V(x_j) \ket{\psi_0}$ to data $x_j$ up to a random initial $\ket{\psi_0} \in \mathcal{H}$. For instance, $V(x) = e^{i \tan^{-1}(x) \s_k}$ injects data $x\in \mathbb{R}$ into a 1-qubit space in a non-linearly fashion~\cite{mitarai2018quantum}.
Typical choices of $V$ include combinations of Hadamard gates, CNOT gates, and gates generated by Pauli matrices $\mathcal{P} = \{I, \s_1, \s_2, \s_3 \}$.
Subsequently, the encoded state $\ket{\psi_{x_j}}$ is deformed by the parameterized gate $U(\theta)$ such that $\ket{\psi_{x_j}} \mapsto U(\theta) \ket{\psi_{x_j}}$, where $\theta$ represents the learnable parameters subject to certain specific optimization routines. It is noted that the deformation ability of VQC majorly comes from $U(\theta)$ where a convention is taking tensor products of the 1-parameter subgroup generated by $\mathcal{P}$,
\begin{equation}\label{E: variational U}
U(\theta) = \prod_{\ell=1}^k \left( \bigotimes_{q=1}^n e^{ -\frac{i}{2} \theta_q^{(\ell)} \s_q^{(\ell)} } \right) \circ \mathcal{C}_{\ell} \in \mathcal{U}(\mathcal{H})
\end{equation}
where $q = 1,\ldots, n$ is the qubit index, $\ell = 1, \ldots, L$ is the index of variational circuit layers up to $L$ with each $\s_q^{(\ell)} \in \mathcal{P}$, $\mathcal{C}_{\ell}$ is the unitary of all other non-parameterized gates such as CNOT gates etc, and $\theta = \{(\theta_1^{(\ell)}, \ldots, \theta_n^{(\ell)})\}_{\ell = 1}^L \in \mathbb{R}^{nL}$ denotes the collection of all variational (learnable) parameters. Often the 1-parameter subgroup $\theta \mapsto e^{ -\frac{i}{2} \theta \s_k}$ generated by $\s_k \in \mathcal{P}$ is denoted as $\{I,  R_x(\theta), R_y(\theta), R_z(\theta)\}$ correspondingly.

\begin{figure}[htbp]
\vskip -0.15in
\begin{center}
\centerline{\includegraphics[width=1\columnwidth]{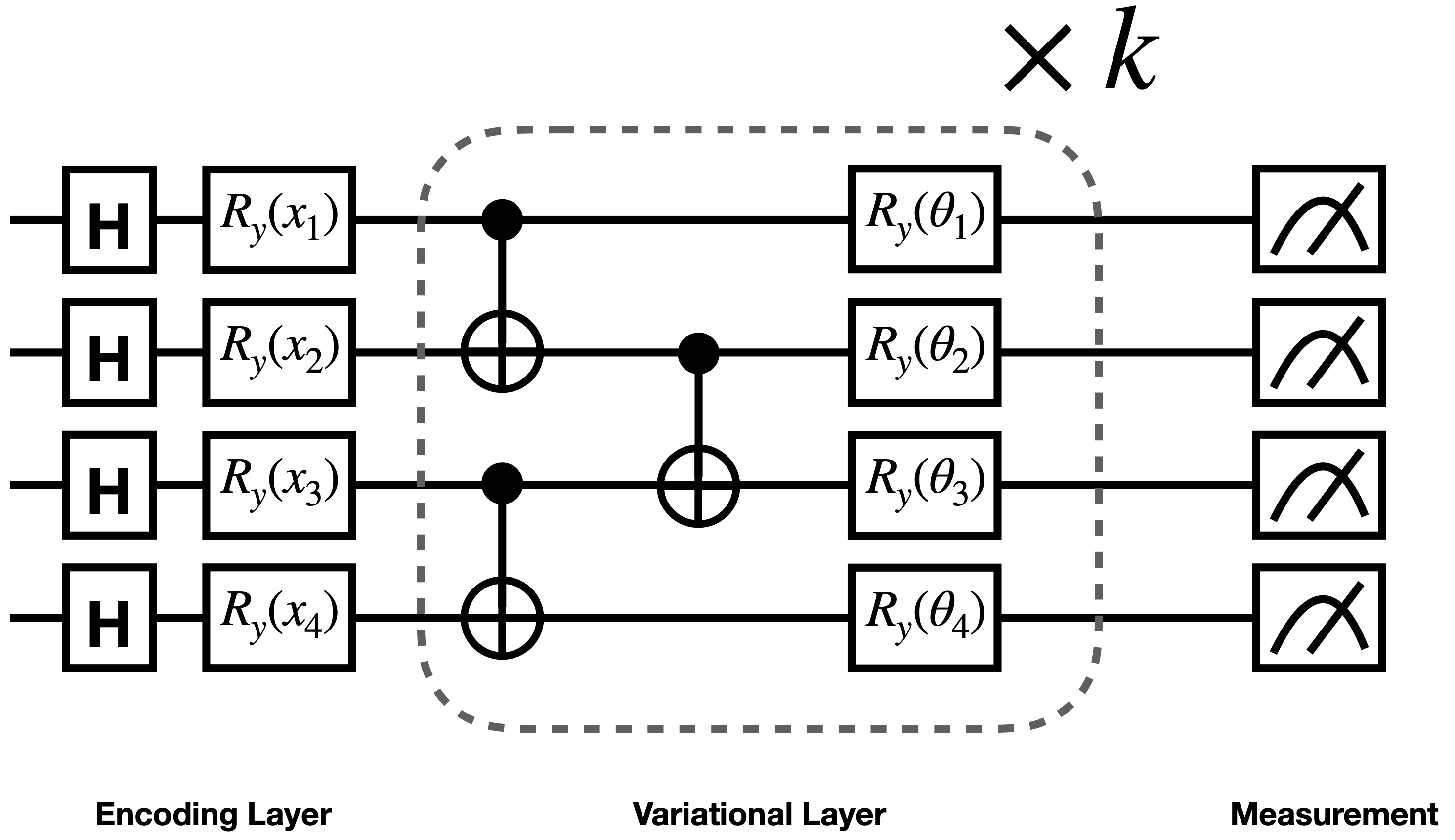}}
\caption{A circuit representation of the VQC used in this work, Eq.~(\ref{E: variational U}). The dashed-line box is repeated $k$ times to increase the depth of the circuit.}
\label{fig: Variational circuits}
\end{center}
\vskip -0.1in
\end{figure}

A final measurement step selects $m$ Hermitian operators $Q_1, \ldots, Q_m $ (each $Q_i \in \mathcal{L}(\mathcal{H})$) to collapse state $U(\theta) \ket{\psi_{x_j}}$ and yields an $m$-dimensional output vector $y = (\langle Q_1 \rangle, \ldots, \langle Q_m \rangle)$ by
\begin{equation}\label{E: quantum expectation}
    \langle Q_i \rangle := \bra{\psi_0} V^{\dagger}(x_j) U^{\dagger}(\theta) \, Q_i \, U(\theta) V(x_j) \ket{\psi_0} \in \mathbb{R}
\end{equation}
with $i=1,\dots, m$. Collectively, the three steps in VQC can be written as $f_{\text{VQC}}:\mathbb{R}^n \to \mathbb{R}^m$ transforming $x_j \mapsto f_{\text{VQC}}(x_j)$, see Fig.~\ref{fig: VQC}. By writing $f_{\text{VQC},\theta}$ we emphasize the dependency on $\theta$. We can drop the subscript $\theta$ when the context is clear. 
Additional components, such as a classical neural network, is possible to be implemented to process raw data into a latent vector of smaller dimensions suitable for a VQC \cite{mari2020transfer,chen2021end,chen2022variational}. This technique is particularly useful when the data dimension exceeds the current capabilities of quantum devices or simulators.
Furthermore, the output values from the VQC $y = (\langle Q_1 \rangle, \ldots, \langle Q_m \rangle)$ can be refined through further quantum or classical processing. For example, the output from a VQC can be processed by another classical neural network or a VQC. This is useful when the desired values are in a range not provided by the quantum observables and require certain rescaling.
A loss function can be chosen as
\begin{equation}\label{E: loss function}
L(F_{\phi}, G_{\eta}, f_{\text{VQC}, \theta}; \mathcal{D}) = \sum_j d \left( y_j, (G_{\eta} \circ f_{\text{VQC}, \theta} \circ F_{\phi})(x_j)\right)
\end{equation}
where $F_{\phi}$ represents the classical pre-processing network, $G_{\eta}$ denotes the classical post-processing network, and $d$ is a function measuring the \emph{distance} between the predicted results and the ground truth $y_{j}$. 
%
The whole hybrid quantum-classical model, including both classical and quantum parameters, can be trained in an end-to-end manner through gradient-based \cite{chen2021end} or gradient-free \cite{chen2022variational} optimization algorithms. The goal is to find the optimal $\Theta^*$, the collection of all quantum and classical trainable parameters $\{\theta, \phi, \eta\}$, by $\Theta^* = \argmin_{\Theta} L(F_{\phi}, G_{\eta},f_{\text{VQC}, \theta}; \mathcal{D})$. 
%

VQC-based models have been proved to be able to outperform classical NN when certain conditions are met~\cite{du2020expressive,abbas2021power}. For example, it is possible to train QML models trained on  smaller training dataset~\cite{caro2022generalization,caro2023out} while maintaining the generalizability. Empirically, VQC has been shown to be successfully in various ML tasks~ \cite{chen2020variational,chen2021end,chen2022quantumCNN,chen2022quantumLSTM,di2022dawn,lockwood2020reinforcement,yun2023quantum,mitarai2018quantum}.

\subsection{\label{sec:Regularities_VQC}Regularities of VQC compared to neural networks}

Fully-connected networks are the building blocks of classical networks which are of the form,
\begin{equation}\label{Eq: fully-connected net}
f = f_n \circ f_{n-1} \circ \cdots \circ f_1 
\end{equation}
where each layer $f_k(z) = \sigma_k (W_k \, z + b_k): \mathbb{R}^{\ell_{k-1}} \to \mathbb{R}^{\ell_k}$ is composed of an activation function $\sigma_k$, a bias vector $b_k \in \mathbb{R}^{\ell_k}$ and a \emph{weight} matrix $W_k \in \mathcal{L}\left( \mathbb{R}^{\ell_{k - 1}}, \mathbb{R}^{\ell_k} \right)$. Here $\mathcal{L}(A, B)$ denotes the collection of linear maps between linear spaces $A$ and $B$. Typically, there is no restriction on (classical network) weights $W_k$ to be trained such that very often $W_k$ is not \emph{invertible} even if $\dim A = \dim B$. This results in a problem where a network prediction $y \in \mathbb{R}^m$ is hard to be traced back as $W_k^{-1}(y)$ does not exist nor $f_k^{-1}(y)$ is well-defined due to the structure of activation functions. Owing to this reason, classical networks are called black boxes and thus lack certain regularity even though the prediction ability is powerful.

On the contrary, since VQCs are comprised of unitary matrices, all gates $U$ are invertible, and a quantum state $\ket{\psi} := U \ket{\psi_0}$ is easily revertible by $\ket{\psi_0} = U^* \ket{\psi}$. From this point of view, VQC possesses certain transparency and better regularity.

Additionally, in view of matrix groups, variational gates $U(\theta)$ in Eq.~(\ref{E: variational U}) form a Lie subgroup of $GL(\mathcal{H})$, all invertible matrices in $\mathcal{L}(\mathcal{H})$, so that $\{ U(\theta) \}$ naturally inherited smooth structures from the differentiable submanifold~\cite{Lee2012}. This provides us some hints that the VQC training may be more stable as the training iterations $\theta^{(\text{iter})} \mapsto U(\theta^{(\text{iter})})$ being contained on the smooth submanifold $\mathcal{U}(\mathcal{H})$. Motivated by these properties, it leads us to consider viewing the explainability of VQC via a classical technique Grad-CAM.


\subsection{\label{sec:QGRAD_CAM_with_VQC}Quantum Grad-CAM by VQC}

Grad-CAM is a technique to interpret and visualize the decisions of CNNs via the gradient calculation of a target classifier. Let a set of CNN filters be $\{\mathcal{W}_{s_1, s_2,c,k} \}_{s_1=1, s_2=1,c=1, k=1}^{S, S, C, K}$ where $C$ is the number of input channels and $K$ is the number of output channels, and $S$ is the kernel size. A \emph{feature map} $\{ A^k_{ij} \}_{i,j,k=1}^{W, H, K}$ is the convolution (output) of the CNN kernels with an input image $x = \{ x_{i, j, c} \}_{c=1}^C$ of $C$ channels given by,
\begin{equation}\label{E: feature activation}
    A^k_{ij} := \sum_{s_1, s_2, c}^{S, S, C} \mathcal{W}_{s_1, s_2, c, k} \cdot x_{i + s_1 - 1, j + s_2 - 1, c} + b_k
\end{equation}
where $W, H \in \mathbb{N}$ are the output image size, $i = 1, \ldots ,W$ and $j = 1, \ldots ,H$ are the output pixel location on the $k^{th}$ channel and $b_k$ is the associated bias. 


It has been recognized that each filter serves a specific purpose to detect certain features, such as edges, textures, or patterns. The convolutional output Eq.~(\ref{E: feature activation}) is also called the \textit{activation map} to emphasize that certain input regions are highlighted and \textit{activated} by the filters.

\begin{figure}[htbp]
\begin{center}
\centerline{\includegraphics[width=1\columnwidth]{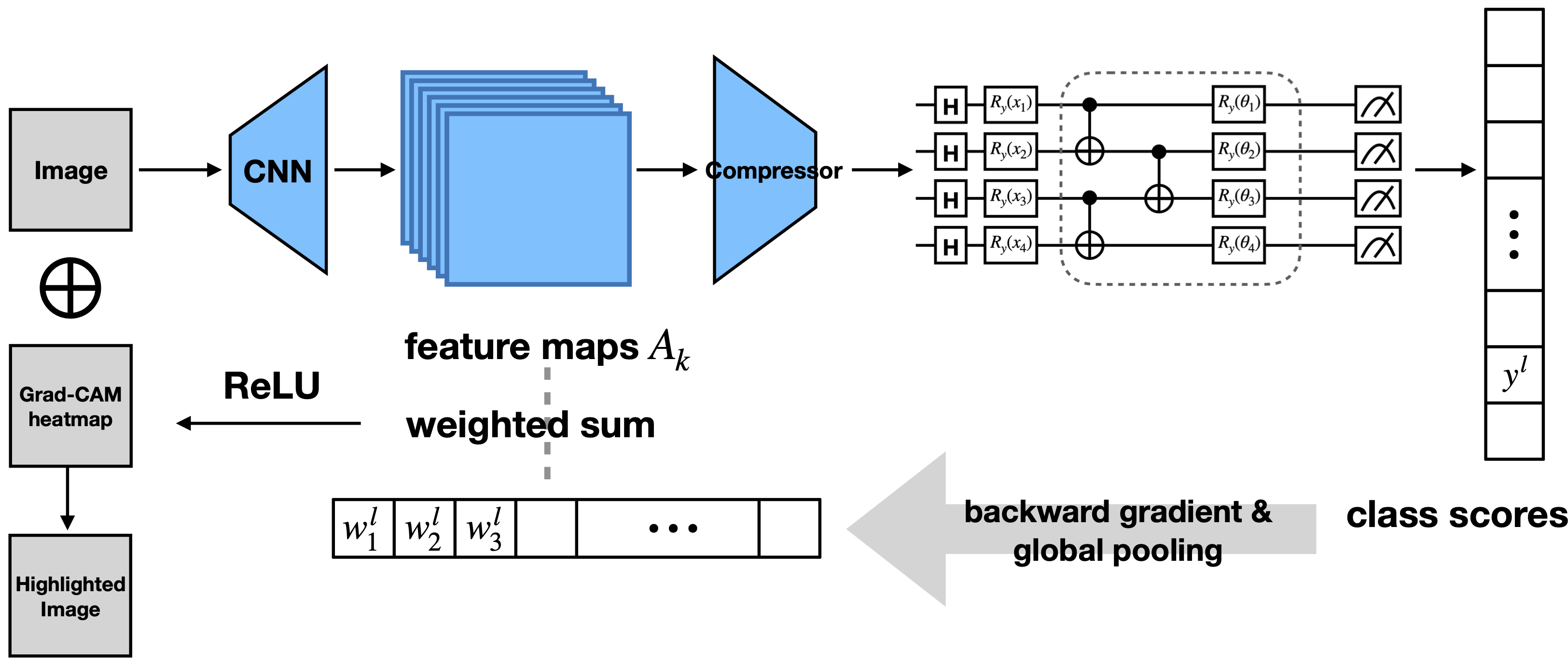}}
\caption{The workflow of Quantum Grad-CAM.}
\label{fig: workflow}
\end{center}
\vskip -0.2in
\end{figure}

Our proposed method, QGrad-CAM is designed to probe the importance of the activation maps $\{ A^k_{ij} \}$ via a VQC classifer with respect to the $K$ filtered channels. Specifically, if a classifier $f:\mathbb{R}^{WHK} \to \mathbb{R}^m$ of $m$ classes is welded after the CNN output $\{ A^k_{ij} \}$ (see Fig.~\ref{fig: workflow}) such that 
\begin{equation}
f(A^k_{ij}) = \left( f^1(A^k_{ij}), \ldots, f^m(A^k_{ij}) \right) \in \mathbb{R}^m
\end{equation}
where each $f^{\ell}(A^k_{ij})$ is the prediction for class $\ell = 1, \ldots, m$, then a \textit{weighting function} $w^{\ell}_k: \mathbb{R}^{WHK} \to \mathbb{R}$ associated to classifier $f$ can be defined,
\begin{equation}\label{Eq: weight}
w_k^{\ell} (A^k_{ij}) := \frac{1}{WH} \sum_{i,j}^{W,H} \frac{\p f^{\ell}(A^k_{ij})}{\p A^k_{ij}},
\end{equation}
with the gradients of the class predictions computed and averaged out. Consequently, the Grad-CAM heatmap is obtained by the composition of the ReLU function with a weighted sum of feature maps Eq.~(\ref{E: feature activation}), (\ref{Eq: weight}),
\begin{equation}
(\text{Grad-CAM heatmap})_{ij} = \text{ReLU} \left( \sum_k^K  A^k_{ij} \cdot w_k^{\ell}(A^k_{ij}) \right),
\end{equation}

The construction of the weighting function Eq.~(\ref{Eq: weight}) naturally inherits important information for the final classification. Thus, it serves as a fundamental indicator for the resulting output.

\section{\label{sec:Explainability_QGRAD_CAM}Explainability by Quantum Grad-CAM}

It is our finding that the importance weighting $w^\ell_{k}$ Eq.~(\ref{Eq: weight}) can be explicitly computed in certain cases of VQC, and the role of each image channel can be understood from the perspective of VQC. In contrast to the classical network, QGrad-CAM gives rise to a certain degree of explainability and learning transparency, which is the key investigation of this study.

Consider the quantum encoding depicted by Fig.~\ref{fig: Variational circuits},
\begin{equation}\label{E: encoding V}
    V(x) = \bigotimes_{q=1}^n e^{ -\frac{i}{2} x_q \, \s_{k_q} } \circ H_q
\end{equation}
where $x = (x_1, \ldots, x_n) \in \mathbb{R}^n$ is an input of VQC, $H_q$ is \emph{any gate} on the $q^{th}$ qubit not related to $x$, such as the Hadamard gate, and $\s_{k_q}$ is a Pauli matrix of index $k_q \in \{0, 1, 2, 3\}$ in $\mathcal{P}$ depending on the $q^{th}$ qubit. Then the measurement of an observable $Q \in \mathcal{L}(\mathcal{H})$ can be computed by a generalized form of Eq.~(\ref{E: quantum expectation}) in terms of a density matrix $\rho \in \mathcal{L}(\mathcal{H})$,
\begin{equation}\label{E: quantum expectation2}
\langle Q \rangle (x) = tr \left( Q \, U(\theta) V(x) \, \rho_0 \, V^{\dagger}(x) U^{\dagger}(\theta) \right)
\end{equation}
where $\rho_0 = \ket{\psi_0} \otimes \bra{\psi_0}$, $U(\vt)$ is as Eq.~(\ref{E: variational U}) and $tr$ is the trace operation on $\mathcal{L}(\mathcal{H})$. We calculate,
\begin{multline}\label{E:dexp}
   \frac{\p \langle Q \rangle }{\p x_q} (x) = tr \left( Q \, U(\theta) \frac{\p V(x)}{\p x_q} \, \rho_0 \, V^{\dagger}(x) U^{\dagger}(\theta) \right) \\ + tr \left( Q \, U(\theta) V(x) \, \rho_0 \, \frac{\p V^{\dagger}(x)}{\p x_q} U^{\dagger}(\theta) \right) 
\end{multline}
Since the differential of a tensor product $x \mapsto A(x) \otimes B(x)$ is defined as,
\begin{equation}
    \frac{\p}{\p x_q} \left( A(x) \otimes B(x) \right) =  \left( \frac{\p A(x)}{\p x_q}  \right) \otimes B(x) +  A(x) \otimes \left( \frac{\p B(x)}{\p x_q} \right)
\end{equation}
we have,
\begin{equation}\label{E: density decomposition}
    \frac{\p}{\p x_q} V(x) = -\frac{i}{2} V_{k_1}(x_1) \otimes \cdots \Bigg( \s_{k_q} \, V_{k_q}(x_q) \Bigg) \cdots \otimes V_{k_n}(x_n)
\end{equation}
where we denote $V_{k_q}(x_q) := e^{ -\frac{i}{2} x_q \, \s_{k_q} } \circ H_q$ for simplicity. We expand a density matrix by the tensorial basis $\{\s_{i_1} \otimes \cdots \otimes \s_{i_n}\}$,
\begin{equation}
    \rho_0 = \sum_{i_1, \ldots,i_n} C_{i_1 \cdots i_n} \, \s_{i_1} \otimes \cdots \otimes \s_{i_n}
\end{equation}
for some coefficients $C_{i_1 \cdots i_n} \in \mathbb{C}$. Then Eq.~(\ref{E:dexp}) yields,
\begin{multline}\label{E:dexp2}
       \frac{\p \langle Q \rangle }{\p x_q} (x) = -\frac{i}{2} \sum_{i_1, \ldots,i_n} C_{i_1 \cdots i_n} \, tr\Bigg\{ U^{\dagger}(\theta) Q U(\theta) \\
       \Big( V_1(x_1) \s_{i_1} V_1^{\dagger}(x_1) \Big) \otimes \cdots \otimes \Big[ \s_{k_q}, V_{k_q}(x_q) \s_{i_q} V_{k_q}^{\dagger}(x_q) \Big] \otimes \cdots \\
       \otimes \Big( V_n(x_n) \s_{i_n} V_n^{\dagger}(x_n) \Big) \Bigg\}
\end{multline}
where the middle term contains a Lie bracket $[A, B]:= AB - BA$ at the $q^{th}$ qubit. In fact, it can be explicitly computed,
\begin{multline}\label{E: parameter shift}
\Big[ \s_{k_q}, V_{k_q}(x_q) \s_{i_q} V_{k_q}^{\dagger}(x_q) \Big] = i \left( V_{k_q} \left(x_q + \frac{\pi}{2}\right) \s_{i_q} V_{k_q}^{\dagger} \left(x_q + \frac{\pi}{2}\right) \right. \\
- \left.  V_{k_q} \left(x_q - \frac{\pi}{2} \right) \s_{i_q} V_{k_q}^{\dagger} \left(x_q - \frac{\pi}{2}\right)  \right)
\end{multline}
Together, Eq.~(\ref{E:dexp2}) and (\ref{E: parameter shift}) give us the explicit formula of the importance weighting Eq.~(\ref{Eq: weight}) in the VQC case. This then reveals how a VQC views the image channels and makes important selections in the sense of Grad-CAM. 

\section{\label{sec:Experiment}Experiment}

QGrad-CAM is applied to three datasets: MNIST, Dogs vs. Cats, and the TIMIT corpus \cite{garofolo1988getting}, each with its respective classification task. Training is conducted end-to-end on all parameters of a 3-layer CNN and a 4-block VQC, both initialized from scratch. The gradient is computed at the end of the VQC towards the last layer of the CNN to generate the results.


\textbf{Image classifications.} \quad  The MNIST dataset consists of grayscale images sized $28 \times 28$, labeled across 10 classes. The Dogs vs. Cats dataset contains color images, resized to $128 \times 128$ for binary classification of dogs and cats.

\begin{figure}
\centering
\includegraphics[width=\columnwidth]{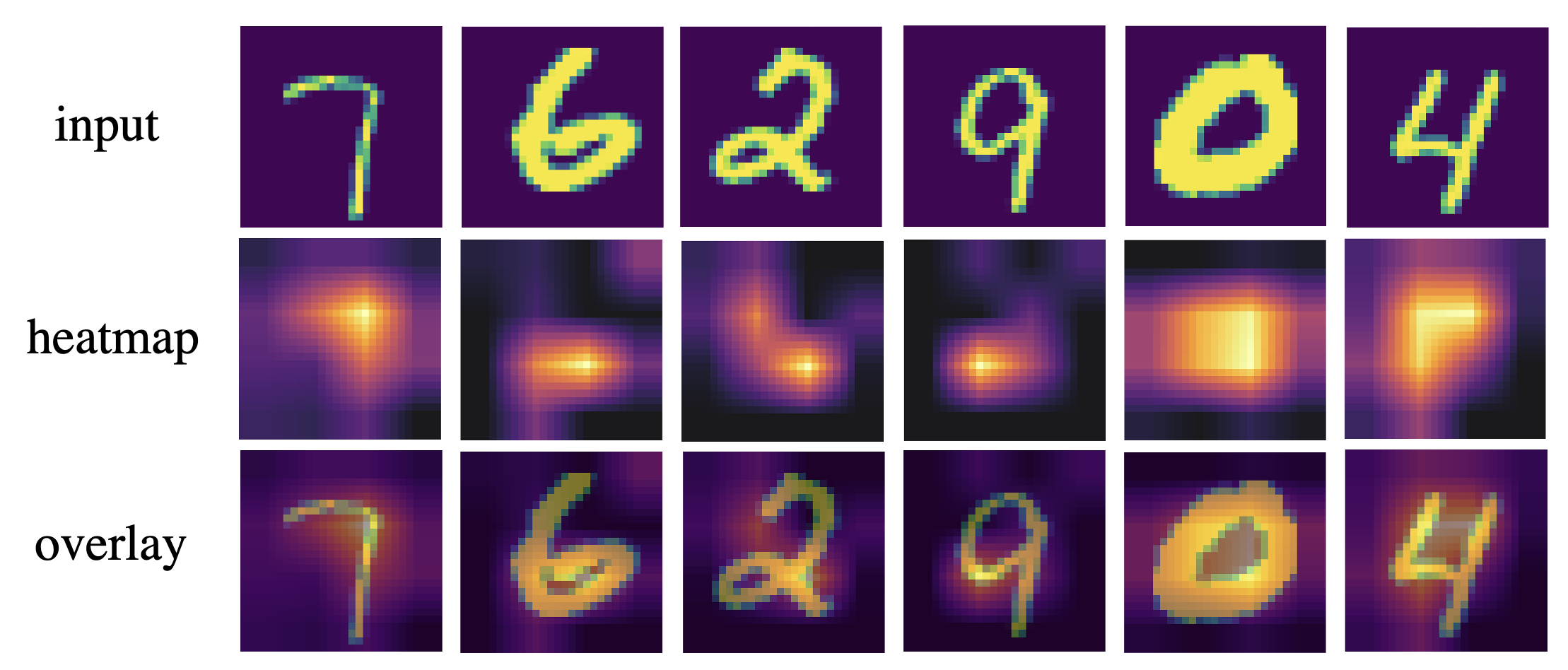}
\caption{\textbf{MNIST.} The generated CAM heatmaps highlight the regions with unique and distinguishable shapes and contours for each digit. For example, the heatmap for the digit `7' emphasizes the sharp turn, while the heatmap for `6' focuses on the closed loop.}
\label{fig: mnist}
\vskip -0.1in
\end{figure}
\begin{figure}
\centering
\includegraphics[width=0.85\columnwidth]{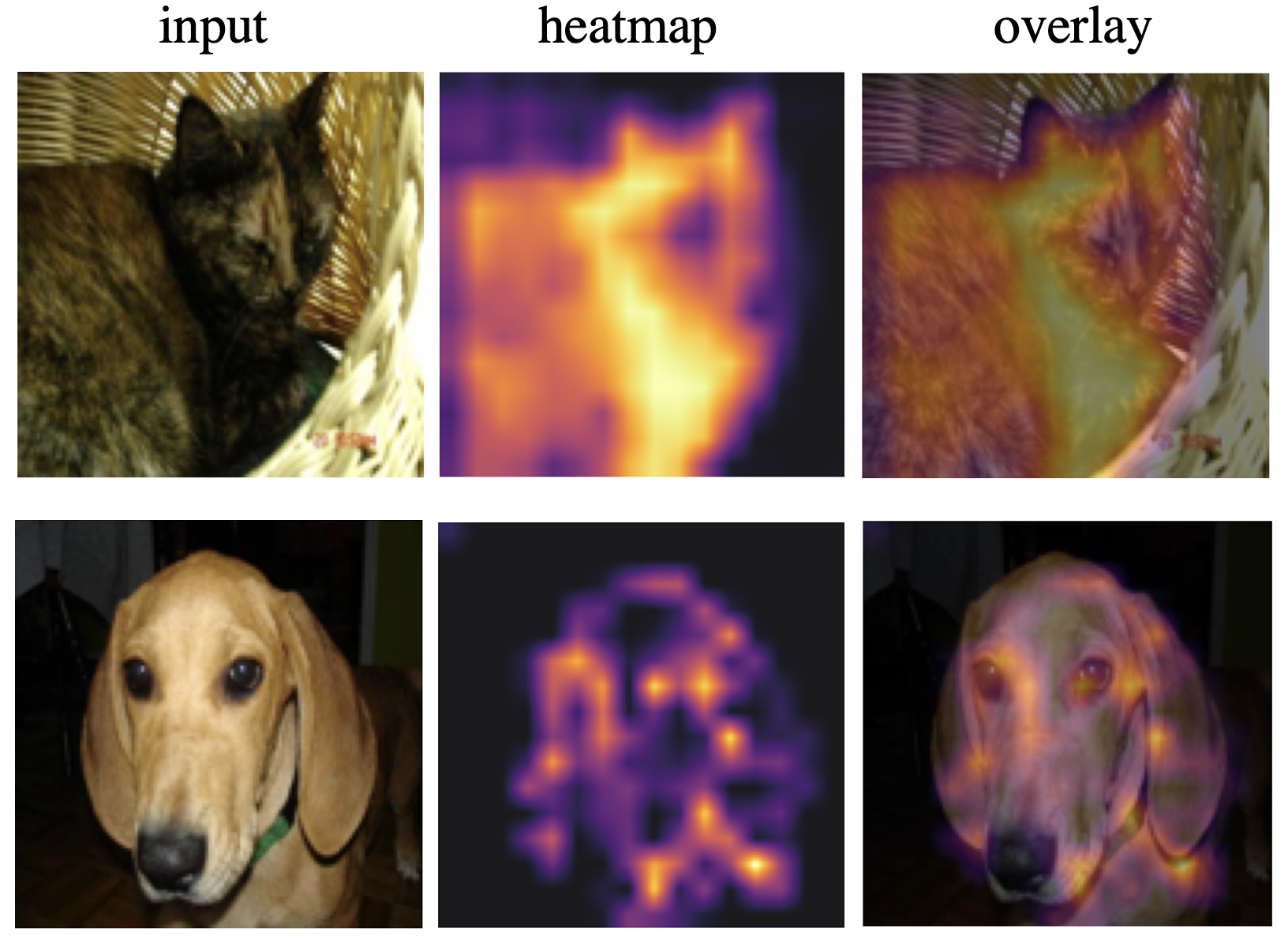}
\caption{\textbf{Dogs vs. Cats.} The higher resolution and colored images allow the detection of fine-grained textures and contours, which are crucial for identifying unique features to each animal.}
\label{fig: catsndogs}
\vskip -0.1in
\end{figure}

Example results of QGrad-CAM with MNIST and Dogs vs. Cats are demonstrated in Fig~\ref{fig: mnist} and Fig~\ref{fig: catsndogs}. It is observed that the generated QGrad-CAMs effectively highlight the regions where key features of each class are located. For MNIST, the heatmaps generally focus on areas where lines turn and curves form, emphasizing angles and the smoothness or sharpness of lines as discriminative features. For the Dogs vs. Cats dataset, the VQC classifier seems to learn critical textures and contours unique to cats and dogs. The heatmaps successfully identify the discriminative regions used for categorization.




\textbf{Speech classifications.}\quad TIMIT contains recordings of American English speakers. The samples are in the 16 kHz WAV format and converted by Short-Time Fourier Transform (STFT) into spectrograms as inputs. Random samples are selected and corrupted by helicopter noise into background to create two classes: with or without a noisy background.

\begin{figure}
\centering
\includegraphics[width=0.95\columnwidth]{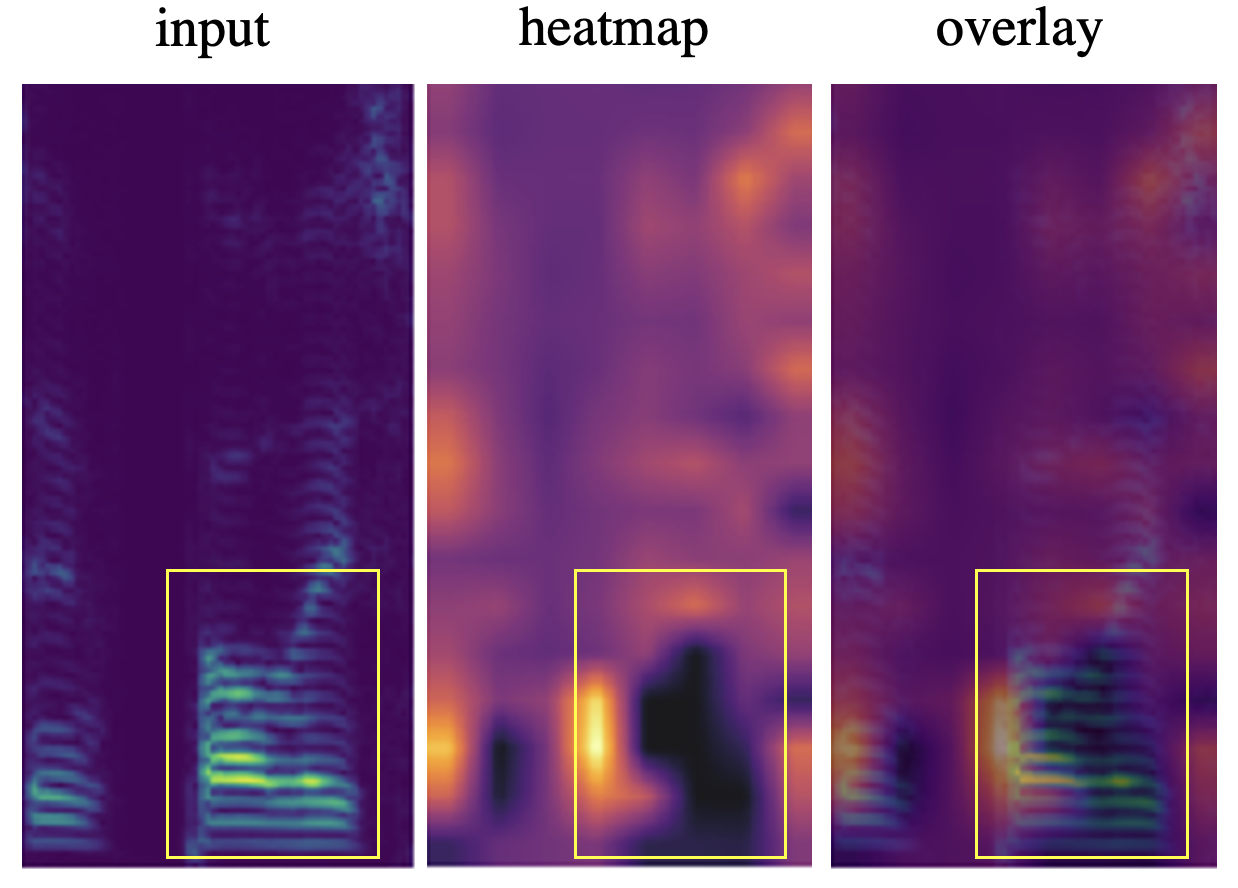}
\caption{\textbf{TIMIT.} Example of the spectrogram of a clean speech utterance. The heatmap has lower intensity inside the area enclosed by the yellow rectangle, where human speech signals are located. }
\label{fig: timit clean}
\vskip -0.1in
\end{figure}
\begin{figure}
\centering
\includegraphics[width=0.95\columnwidth]{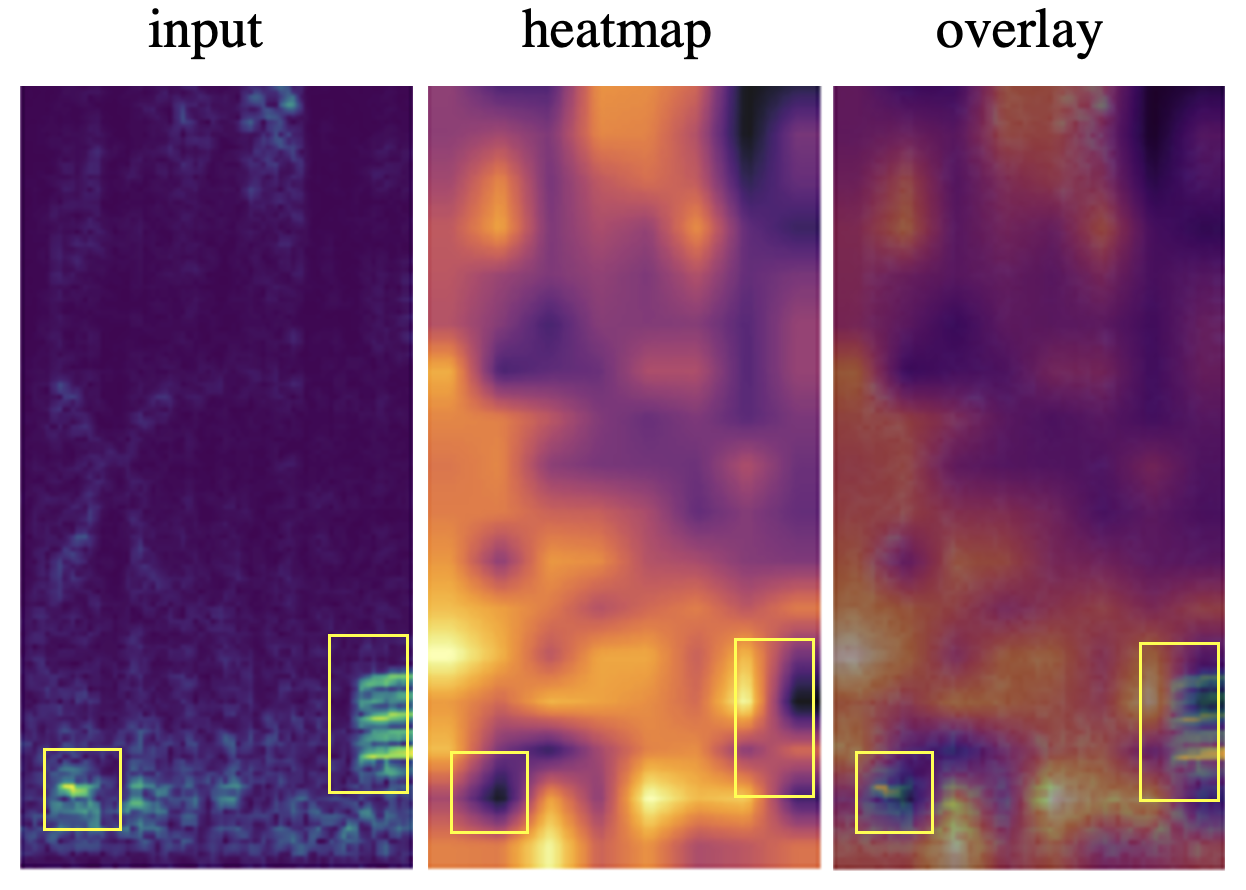}
\caption{\textbf{TIMIT.} Example of a spectrogram for a speech utterance corrupted with helicopter noise. The helicopter noise appears as diffuse areas in the spectrogram. The accompanying heatmap visually indicates that the network focuses on the background, outside the region of the human speech signal.}
\label{fig: timit noisy}
\vskip -0.1in
\end{figure}

A spectrogram provides a visual representation of the frequency content of a speech signal over time. The horizontal axis represents time, the vertical axis represents frequency, and the color intensity indicates the amplitude at each frequency-time point. Human speech typically occupies specific frequency ranges in the spectrogram corresponding to the characteristics of the human vocal tract. In contrast, background noise often appears as diffuse, spread-out patterns across various frequencies and times without distinct features. Figures~\ref{fig: timit clean} and \ref{fig: timit noisy} show spectrograms of clean and noisy speech samples, respectively. Yellow rectangles highlight the regions containing human speech signals for better clarity.


Our results indicate that the network focuses on areas outside the speech utterance to determine whether an utterance is corrupted by noise. This is observed in Figures~\ref{fig: timit clean} and \ref{fig: timit noisy}, where the heatmaps show lower intensity within the yellow rectangles. Rather than concentrating on the portions of the spectrogram containing the speech signal, the network examines the background areas to detect signs of noise, allowing for more accurate identification of corruption.


\section{\label{sec:Conclusion}Conclusion}


Motivated by the structured nature of the quantum framework, this work introduces QGrad-CAM, a novel method for providing visual explanations for model decisions. Our approach integrates the VQC with CNN gradient techniques to generate detailed, class-specific image localization. Experimental results on both image and speech datasets demonstrate the method's effectiveness in highlighting discriminative features. Furthermore, an explicit importance weighting function associated to a VQC classifier can be analytically derived. Our results suggest potential advantages of quantum techniques in enhancing interpretability, highlighting the need for further exploration into the quantum advantage in this area.




\bibliographystyle{IEEEtran}
\bibliography{references,bib/qml_examples,bib/vqc,bib/explain_qml}

\end{document}